\documentclass[aps,prd,preprintnumbers,amsmath,amssymb]
{revtex4}

\input def.sty

\usepackage{epsfig}

\begin{document}

\title{
First Order QED Corrections to the Parity-Violating Asymmetry in
\Moller\ Scattering
}

\affiliation{Gomel State Technical University, Gomel, Belarus}
\affiliation{National Center for Particle and High Energy Physics, Minsk, Belarus}
\affiliation{University of California, Berkeley, California, USA}

\author{\firstname{Vladimir A.}
\surname{Zykunov}}
\affiliation{%
Gomel State Technical University, Gomel, Belarus}

\author{\firstname{Juan}
\surname{Suarez}}
\affiliation{National Center for Particle and High Energy Physics, Minsk, Belarus}

\author{\firstname{Brock A.}
\surname{Tweedie}}
\affiliation{University of California, Berkeley, California, USA}

\author{\firstname{Yury G.}
\surname{Kolomensky}}
\affiliation{University of California, Berkeley, California, USA}

\begin{abstract}
We compute a full set of the first order QED corrections to the
parity-violating observables in polarized \Moller\ scattering. 
We employ a covariant method of removing infrared divergences, 
computing corrections without introducing any unphysical
parameters. 
When applied to the kinematics of the SLAC E158 experiment,
the QED corrections reduce the parity violating asymmetry
by $4.5\%$. We combine our results with the previous calculations of
the first-order electroweak corrections and obtain the complete
$\mathcal(O)(\alpha)$ prescription for relating the experimental asymmetry
\ALR\ to the low-energy value of the weak mixing angle \sinTW. Our results are
applicable to the recent measurement of \ALR\ by the SLAC E158
collaboration, as well as to the future parity violation
experiments. 
\end{abstract}

\maketitle

\section{Introduction}
Measurements of the parity-violating observables in electron
scattering provide information about low-energy structure of 
weak neutral current processes. Such observables arise
from the interference between the weak and electromagnetic
amplitudes~\cite{ref:Zeldovich}, and are sensitive to the electroweak
couplings. In the Standard
Model, the couplings of the $Z^0$ boson to the fermions are determined by
the weak mixing angle $\theta_{\rm W}$, which has been measured with
high precision at the $Z^0$
resonance~\cite{ref:PDG2004}. Precision measurements of 
the low-energy parity-violating observables provide an independent
determination of the weak mixing angle, directly test higher order
electroweak corrections, and provide strong independent constraints on the new
physics contributions at the TeV scales~\cite{ref:MarcianoRosner}.  

The experiment E158 at the Stanford Linear Accelerator Center (SLAC)
has measured the parity-violating left-right asymmetry \ALR\ in \Moller\
scattering of polarized 50~GeV electrons off unpolarized atomic
electrons in a liquid hydrogen target~\cite{ref:E158}. The final
uncertainty on \ALR\ is about 10\%.
The measurement of \ALR\ translates into
the measurement of \sinTW\ with a precision of
$\sigma(\sinTW)\approx 0.001$ at low momentum transfer 
$Q^2\approx 0.03~{\rm GeV}^2/c^2$, and is sensitive to both
electroweak one-loop radiative corrections and new physics
phenomena at the TeV scales. 

A precise comparison of the experimental results with the Standard
Model predictions requires detailed understanding of the 
radiative corrections, including effects of the QED bremsstrahlung.
The leading-order
electroweak corrections to the \Moller\ scattering have been computed
a number of years ago~\cite{cz-marc,denn}. The authors of Ref.~\cite{cz-marc}
factorized out the 
soft bremsstrahlung contribution, but did not include the effects of
hard bremsstrahlung, arguing that their effects are small, and
would require the knowledge of experimental
kinematics and acceptance. 
In this paper, improving on the calculation of Ref.~\cite{zyk},
we present a complete
calculation of the first order QED corrections. The detailed analysis
shows that the QED corrections are indeed small but 
not insignificant, compared to the systematic uncertainties of
E158 and the projected uncertainties of the future parity violation
experiments.

Experimentally, \Moller\ scattering is often used to measure 
polarization of electron beams~\cite{ref:polarimetry}. In such
measurements, both beam and 
target electrons are polarized, and electroweak effects can typically
be neglected. QED corrections to the
parity-conserving polarized \Moller\ scattering have been computed in
Ref.~\cite{suarez,IlyichevZykunov} and are relatively important,
compared to the 
typical precision of \Moller\ polarimeters. Similar to
Ref.~\cite{suarez,IlyichevZykunov}, we perform our calculations in the
covariant 
framework of Bardin and Shumeiko~\cite{covar}, which allows us to cancel out
infrared divergences without introducing unphysical parameters (such
as a frame-dependent cutoff $\Delta E$ that separates the soft
bremsstrahlung region from the hard bremsstrahlung contributions). 

This paper is organized as follows. \secref{sec:born} introduces the
kinematics of the \Moller\ scattering and Born cross section and
parity-violating asymmetry. \secref{sec:ir} explains the
regularization of infrared divergences. \secref{sec:results} presents
numerical results applied to the kinematics of SLAC E158
experiment. 

\section{Definitions, Born Cross Section, and Kinematics}
\label{sec:born}

\begin{figure}
\epsfig{file=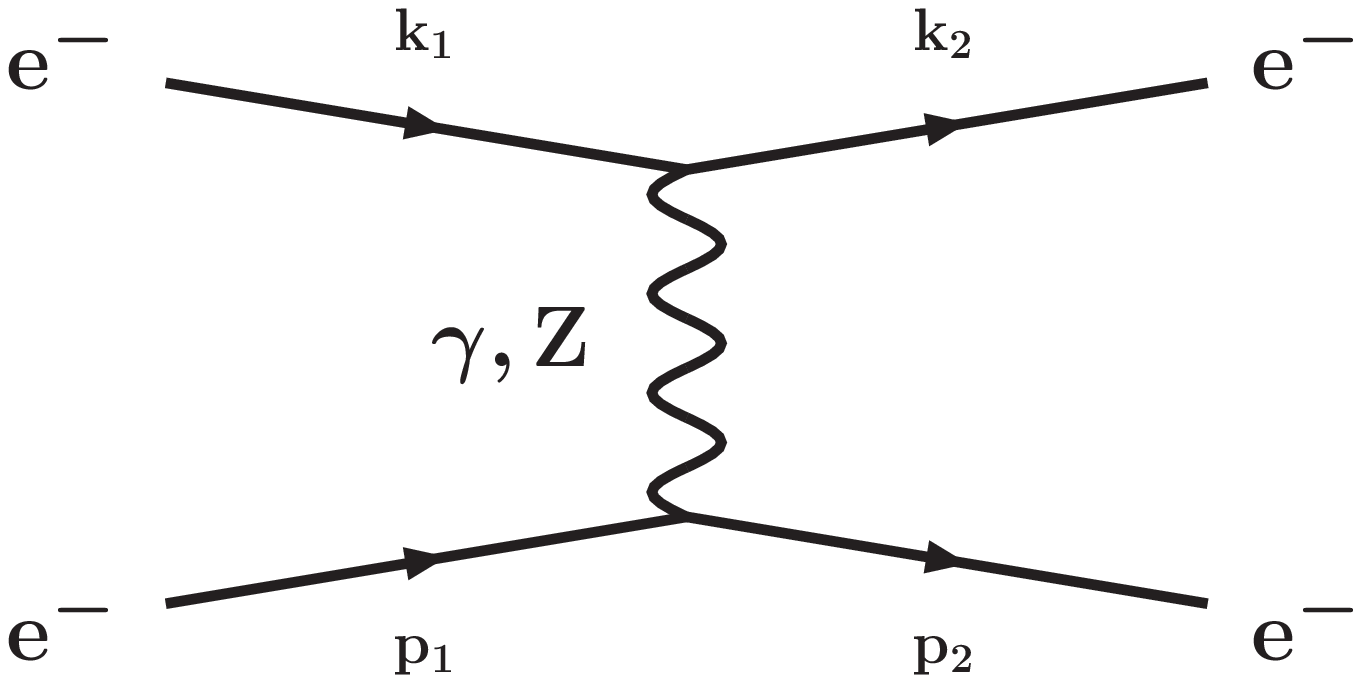,width=2.5in} 
\epsfig{file=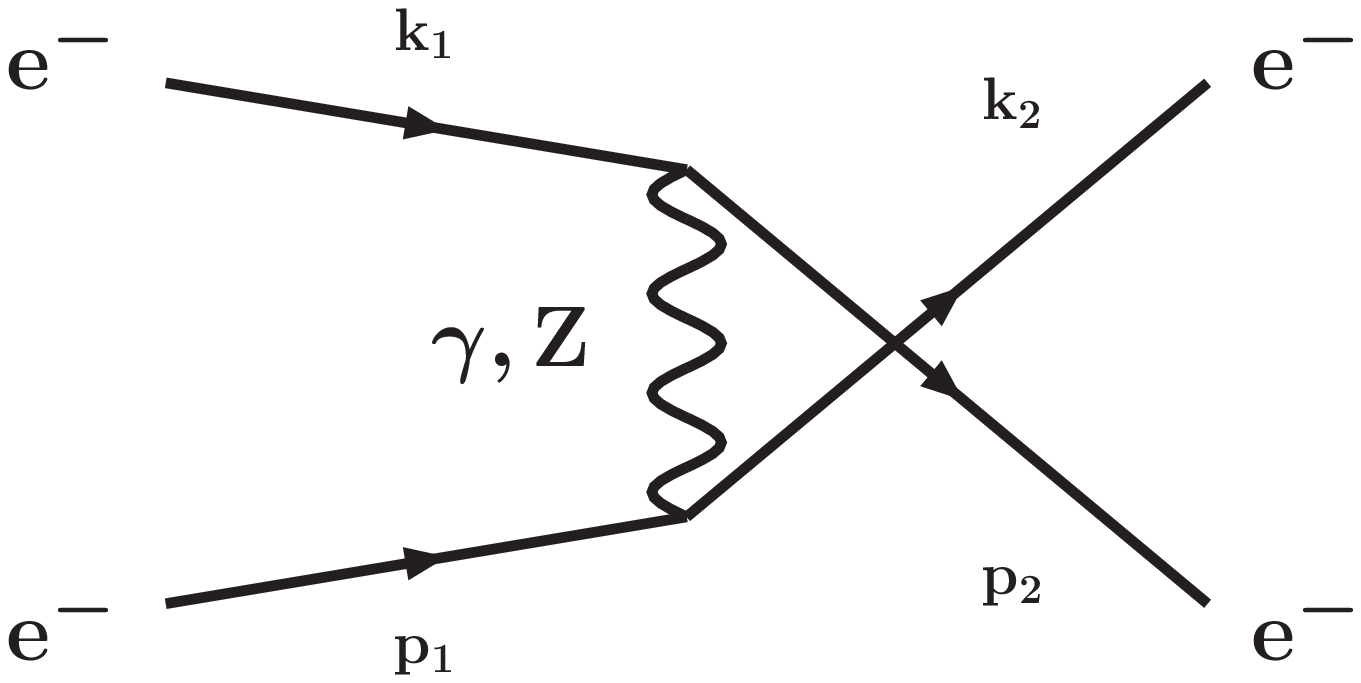,width=2.5in} 
\caption{
Born-level diagrams for \Moller\ scattering: $t$-channel (left) and
$u$-channel (right).
}
\label{born}
\end{figure}

The lowest-order (Born) cross section for M{\o}ller scattering,
defined by the Feynman diagrams in \figref{born}, can be written as
\begin{widetext}
\beq
\sigma^0 =\frac{2\pi \alpha^2}{s}
\sum_{i,j=\gamma,Z} [\lambda_-^{ij}(u^2D^{i}(t)D^{j}(t)+t^2D^{i}(u)D^{j}(u))
   + \lambda_+^{ij}s^2(D^{i}(t)+D^{i}(t))(D^{j}(t)+D^{j}(u))],
\label{cs0} 
\eeq
\end{widetext}
where the four-momenta of the initial and final electrons
$k_1,\;p_1$ and $k_2,\;p_2$ (see Fig.\ref{born}) are combined
to form Mandelstam invariants
\begin{equation}
s=(k_1+p_1)^2,\ t=(k_1-k_2)^2,\ u=(k_2-p_1)^2\ .
\end{equation}
In Eq.~(\ref{cs0}) and hereafter, we use a short-hand notation $\sigma
\equiv d\sigma/dy$. 

The Born cross section in Eq.~(\ref{cs0}) is written in terms of the
photon and $Z^0$ propagators
\begin{equation}
D^{i}(k) \equiv D^{ik} =\frac{1}{k-m_i^2}\ (i=\gamma,Z)
\end{equation}
and the coupling factors
\beq
\lambda_{\pm}^{ij} = {\lambda_1}_B^{ij}{\lambda_1}_T^{ij}
                       \pm {\lambda_2}_B^{ij}{\lambda_2}_T^{ij}\ .
\eeq
The latter in turn depend on the polarizations of the beam ($p_B$)
and target ($p_T$) electrons:
\begin{equation}
 {\lambda_1}_{B(T)}^{ij} = \lambda_V^{ij} -p_{B(T)} \lambda_A^{ij}\ ,\ 
   {\lambda_2}_{B(T)}^{ij} = \lambda_A^{ij} -p_{B(T)} \lambda_V^{ij},
\end{equation}
\begin{equation}
 \lambda_V^{ij}=v^iv^j + a^ia^j,\
   \lambda_A^{ij}=v^ia^j + a^iv^j,
\end{equation}
where $v^{i},\ a^{i}$ are the vector and axial-vector coupling
constants for the photon and $Z^0$:
\begin{equation}
v^{\gamma}=1,\ a^{\gamma}=0,
\end{equation}
\begin{eqnarray}
  v^Z&=&(-1+4s_W^2)/(4s_Wc_W) \nonumber \\
  a^Z&=&-1/(4s_Wc_W)
\end{eqnarray}
and $s_W$ ($c_W$) is sine (cosine) of the Weinberg angle.

It is convenient to rewrite the cross section in terms of four
Born-level matrix elements $M^0_l$:
\beq
\sigma^0=\frac{\pi \alpha^2}{s}
\mathrm{Re} \sum_{l=1}^4 ( M^0_l + {\hat M}^0_l ),
\eeq
where matrix elements $\hat{M}^0_l$ are obtained from $M^0_l$ by
crossing symmetry $t\leftrightarrow u$. The matrix elements $M^0_l$
are expressed through the ``even'' and ``odd'' functions $M_{e}$ and
$M_{o}$:
\begin{eqnarray}
\displaystyle
   M^0_1&=&
D^{\gamma t}( D^{\gamma t}M_{e}^{\gamma \gamma \gamma \gamma} -
              D^{\gamma u}M_{o}^{\gamma \gamma \gamma \gamma} ),\
\nonumber \\
   M^0_2&=&
D^{\gamma t} ( D^{Zt}      M_{e}^{\gamma Z\gamma Z} -
               D^{Zu}      M_{o}^{\gamma Z\gamma Z} ),\
\nonumber \\
   M^0_3&=&
D^{Zt} ( D^{\gamma t}      M_{e}^{Z\gamma Z\gamma} -
         D^{\gamma u}      M_{o}^{Z\gamma Z\gamma}  ) ,\
\nonumber \\
   M^0_4&=&
D^{Zt} ( D^{Zt}M_{e}^{ZZZZ} -
         D^{Zu}            M_{o}^{ZZZZ} )\ .
\end{eqnarray}
The matrix elements $M_{e}$ and $M_{o}$ are defined in such a way
that they can be used, with minimal modifications, in both Born
and first-order matrix elements. They are defined in terms of
the couplings $\lambda$
\begin{eqnarray}
M^{ijkl}_{e} &=&
        2 (s^2+u^2){\lambda_1}_B^{ij}{\lambda_1}_T^{kl}
       +2 (s^2-u^2){\lambda_2}_B^{ij}{\lambda_2}_T^{kl},
\\
M^{ijkl}_{o} &=&
       - 2 s^2 ( {\lambda_1}_B^{ij}{\lambda_1}_T^{kl}
                +{\lambda_2}_B^{ij}{\lambda_2}_T^{kl} ) \ .
\end{eqnarray}

The kinematic variable $y$ is defined as
\begin{equation}
y=-\frac{t}{s} \approx \frac{1-\cos \Theta}{2}\frac{E^{*'}}{E^{*}},
\end{equation}
where $\Theta$ is the center of mass (CMS) scattering angle of the
detected electron with momentum $k_2$.
$E^{*}(E^{*'})$ is the energy of the initial (detected) electron in CMS,
respectively.
In the Born approximation, the \Moller\ scattering is elastic and 
$E^{*}=E^{*'}$, therefore 
\beq
y_{\rm Born}=\frac{1-\cos \Theta}{2} = 1-\frac{E^{'}}{E}\ 
\eeq
where $E$ and $E^{'}$ are the initial and scattered electron energies
in the Lab frame, respectively. 
Whenever possible, we ignore the electron mass $m$ (which cannot be
done in the collinear singularity regions).

At the Born level, the unpolarized (averaged over helicity states)
cross section is given analytically by~\cite{Moller}  
\beq
\sigma^0 =
\frac{2\pi\alpha^2}{s y^2(1-y)^2}\left(1+y^4+(1-y)^4\right) \, .
\label{eq:xsect}
\eeq
Polarization asymmetry \ALR\ is conventionally defined as
\begin{equation}
A_{LR} \equiv
 \frac{\sigma_{LL}+\sigma_{LR}-\sigma_{RL}-\sigma_{RR}}
      {\sigma_{LL}+\sigma_{LR}+\sigma_{RL}+\sigma_{RR}}\ ,
\label{A}
\end{equation}
where the first helicity index refers to the beam electrons, and the
second helicity index corresponds to the target electrons.  
Since the
target helicity is summed over, \ALR\ defined by Eq.~(\ref{A}) is a
parity-violating observable\footnote{Notice that \ALR\
has the opposite sign compared to the asymmetry $A_{\rm PV}$ defined
in many low-energy experiments, e.g. Ref.~\cite{ref:E158}. In our
definition, \ALR\ in \Moller\ scattering is positive.}. The Born-level
asymmetry is given by~\cite{cz-marc}
\bea
\ALR^0 &=& \mathcal{A}^0(Q^2,y) \left(1-4\sin^2\theta_W\right) 
=
\frac{G_F s}{\sqrt{2}\pi\alpha}\frac{y(1-y)}{1+y^4+(1-y)^4} \left(1-4\sin^2\theta_W\right)\, ,
\label{eq:ALRtree}
\eea
where $\mathcal{A}^0(Q^2,y)$ is an experimental acceptance-dependent
analyzing power. 
Kinematics of the E158 experiment correspond to the laboratory beam
energies of $E=45 (48)$ GeV and CMS scattering angles
$-0.5<\cos\Theta\le0$, or 
average invariants $s = 2mE \approx 0.048~{\rm GeV}^2/c^2$, 
$Q^2 = -t = 0.026~{\rm GeV}^2/c^2$ and 
$y \approx 0.6$~\cite{ref:E158}.

We perform our calculation in the on-shell (OS)
renormalization scheme, defining the weak mixing angle to all orders
in perturbation theory as $c_W \equiv m_W/m_Z$, where $m_W$ and $m_Z$
are the physical masses of the $W^{\pm}$ and 
$Z^0$ bosons, respectively. For consistency with the precision
electroweak measurements~\cite{ref:PDG2004,ref:EWWG}, we use 
\bea
m_W &=& 80.390~{\rm GeV} \\
m_Z &=& 91.188~{\rm GeV} 
\eea
which implies
\beq
\sin^2\theta_W = 0.2228
\eeq
in the on-shell scheme. As we will note in \secref{sec:results}, while
the absolute value of the Born asymmetry depends on the
electron neutral current coupling $1-4\sinTW$, the {\em relative\/}
corrections to the experimentally measured asymmetries computed here
are quite insensitive to the 
choice of couplings or the renormalization scheme. 

\section{Radiative Corrections}
\label{sec:ir}

It is well known that effects of ``internal'' bremsstrahlung (real
photon emission) need to be
combined with the contributions from the other one-loop (leading order)
electroweak radiative corrections (so-called virtual, or
V-contributions). The virtual contributions to \Moller\ scattering
have been studied extensively~\cite{cz-marc,denn} and are not repeated
here. However, we need the infrared-divergent part of the
V-contributions, which cancels out the corresponding divergences in the
bremsstrahlung cross section. Naturally, the extraction of the
IR-divergent parts is a somewhat subjective procedure, which may lead
to ambiguities similar to the concept of scheme dependence in
the ultraviolet renormalization. We follow the framework of Bardin and
Shumeiko~\cite{covar} and extract the infrared-divergent
contributions that are strictly proportional to the Born cross
section. Such contributions cancel in the parity-violating asymmetry,
and remain small even after other corrections are taken into account. 
We take the infrared-divergent parts of the virtual corrections from
\cite{BSH86,Hol90}, which include the vacuum polarization and
vertex correction contributions. We also compute the IR-divergent
$\gamma\gamma$ and $\gamma Z$ box diagrams.

\begin{figure}
\begin{tabular}{ccc}
\epsfig{file=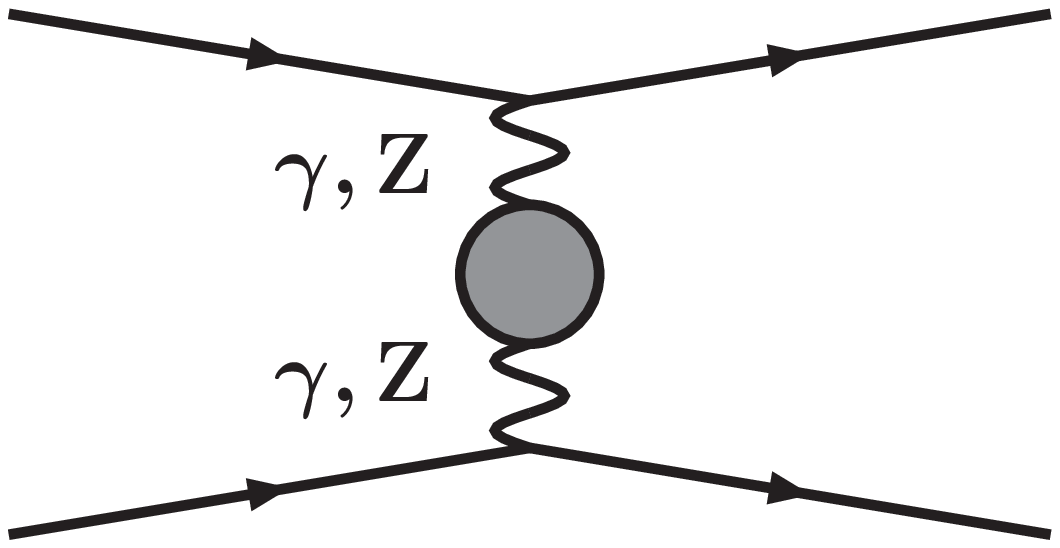,width=2in} &
\epsfig{file=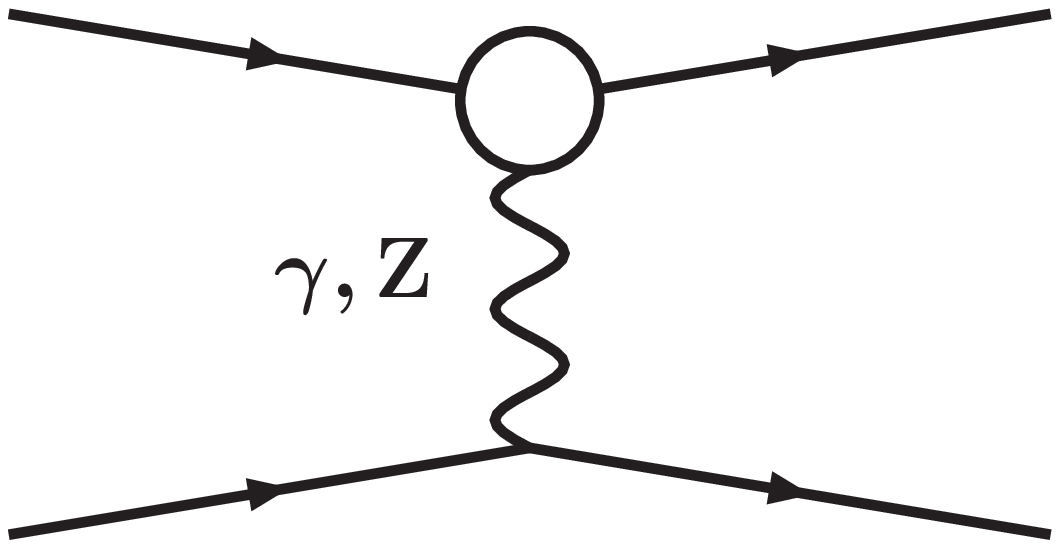,width=2in} &
\epsfig{file=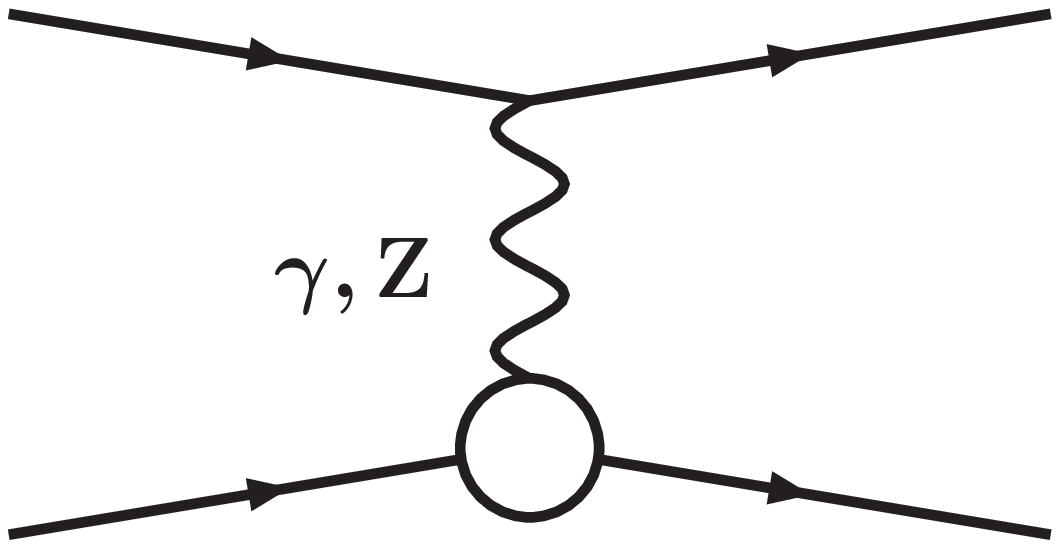,width=2in} \\
(a) & (b) & (c) \\
\epsfig{file=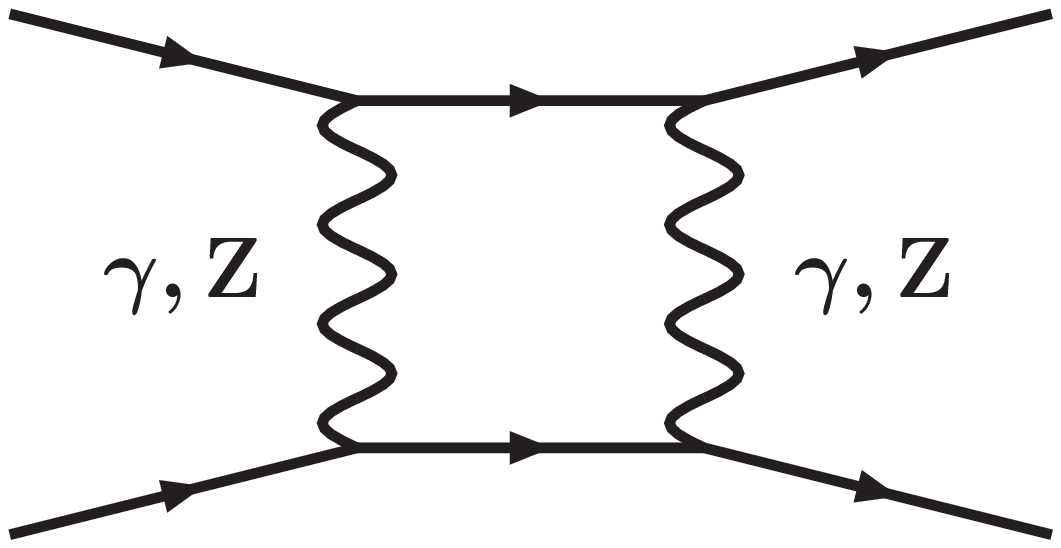,width=2in} &
\epsfig{file=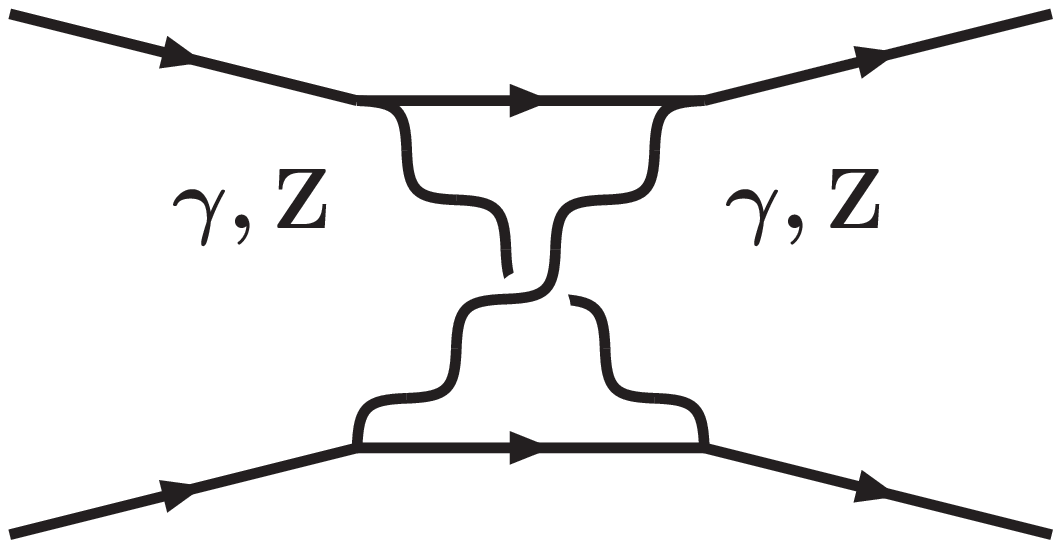,width=2in} &
\epsfig{file=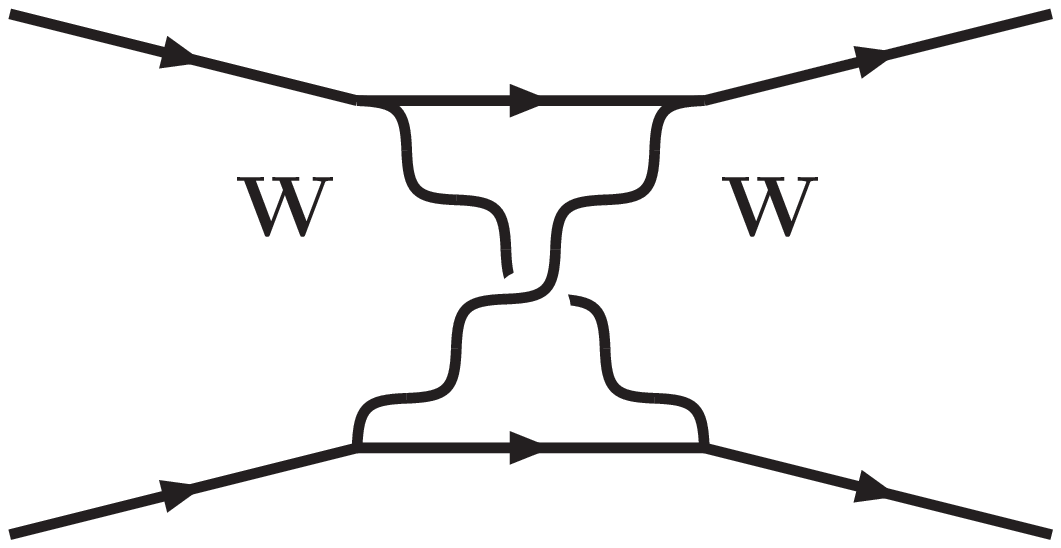,width=2in} \\
(d) & (e) & (f) \\
\end{tabular}
\caption{
One-loop $t$-channel contributions to \Moller\ scattering. Circles
represent vacuum polarization (a) and vertex correction contributions
(b,c), given in \cite{BSH86,Hol90}. 
}
\label{fig:2f}
\end{figure}

The virtual contributions to the \Moller\ scattering cross section and
asymmetry can be classified into three categories: the vacuum
polarizations of the gauge boson propagators, vertex corrections, and
box diagrams (see \figref{fig:2f}):
\begin{equation}
\sigma^{V}=
\sigma^S+\sigma^{Ver}+\sigma^B\ .
\label{G}
\end{equation}

The vacuum polarizations of $\gamma$ and $Z^0$ bosons are
infrared-convergent, and are not considered in this paper. The
infrared-divergent parts of the vertex corrections (Figures
\ref{fig:2f}b and \ref{fig:2f}c) are obtained from form-factors 
$\delta F_{V,A}^{je}$ given in \cite{BSH86} (for $k^2 = t,u$).
Substituting the coupling constants for the vertex form-factors (\eg\ 
$v^{\gamma} \rightarrow \delta F_V^{\gamma e}$) in the expressions for
the Born functions  $M_{e,o}$, we get the vertex part of the cross
section 
\begin{equation}
\sigma^{Ver}=\frac{2\pi \alpha^2}{s}
\mathrm{Re} \sum_{l=1}^4 ( M^V_l + {\hat M}^V_l ),
\label{V}
\end{equation}
where 
\begin{widetext}
\begin{eqnarray}
\displaystyle
M^V_1&=&D^{\gamma t} ( D^{\gamma t}
            ( M_{e}^{ F^{\gamma}\gamma\gamma\gamma}
             +M_{e}^{\gamma\gamma F^{\gamma}\gamma} )
         -D^{\gamma u}
            ( M_{o}^{ F^{\gamma}\gamma \gamma \gamma}
             +M_{o}^{ \gamma\gamma F^{\gamma}\gamma} )),
\nonumber \\
M^V_2&=&D^{\gamma t} ( D^{Z t}
            ( M_{e}^{F^{\gamma}Z\gamma Z}
             +M_{e}^{\gamma ZF^{\gamma}Z} )
         -D^{Z u}
            ( M_{o}^{F^{\gamma}Z\gamma Z}
             +M_{o}^{\gamma ZF^{\gamma} Z} )),
\nonumber \\
M^V_3&=&D^{Z t} ( D^{\gamma t}
            ( M_{e}^{F^{Z}\gamma Z\gamma}
             +M_{e}^{Z\gamma F^{Z}\gamma} )
         -D^{\gamma u}
            ( M_{o}^{F^{Z}\gamma Z \gamma }
             +M_{o}^{Z\gamma F^Z \gamma } )),
\nonumber \\ 
M^V_4&=&D^{Z t} ( D^{Z t}
            ( M_{e}^{F^{Z}ZZZ}
             +M_{e}^{ZZ F^{Z}Z} )
         - D^{Z u}
            ( M_{o}^{F^{Z}ZZZ}
             +M_{o}^{ZZF^{Z}Z} )).
\end{eqnarray}
\end{widetext}

The box diagrams with at least one photon (\ie\ Figures~\ref{fig:2f}d
and \ref{fig:2f}e plus $u$-channel graphs)
also contain infrared divergences. The
diagrams with two $Z^0$ or two $W$ bosons are infrared-convergent
and are not considered here. 
We compute the box diagram contribution as a sum of the
infrared-divergent and infrared-finite parts 
$\sigma^B=\sigma^B_{F}+\sigma^B_{IR}$.
The IR-finite parts of the $\gamma\gamma$ and $\gamma Z$
boxes are expressed by
\begin{equation}
\sigma^B_{F}=-\frac{2\alpha^3}{s}
\sum_{k=\gamma,Z} \left( B_{\gamma\gamma}^k + B_{\gamma Z}^k \right)
+ \bigl(  t \leftrightarrow u  \bigr)\ .
\end{equation}
The terms $B$ have the following form:
\begin{eqnarray}
B_{(\gamma \gamma)}^k&=&D^{kt}
\lambda_-^{\gamma k} \delta^1_{(\gamma \gamma)} +
(D^{kt}+D^{ku})\lambda_+^{\gamma k} \delta^2_{(\gamma \gamma)}
\nonumber \\
B_{(\gamma Z)}^k&=&D^{kt}
\lambda_-^{Z k} \delta^1_{(\gamma Z)} +
(D^{kt}+D^{ku})\lambda_+^{Z k} \delta^2_{(\gamma Z)}
\end{eqnarray}

At low energies ($s, |t|, |u|\; \ll \; m^{2}_{Z}$),
$\delta_{(ij)}^{1,2}$ are given by the following 
expressions: 
\begin{eqnarray}
 \delta^1_{(\gamma \gamma)} & = &
  L_s^2\frac{s^2+u^2}{2t} - L_su -(L_x^2+\pi^2)\frac{u^2}{t}
\nonumber \\
     \delta^2_{(\gamma \gamma)} &=&
  L_s^2\frac{s^2}{t} + L_x s - (L_x^2 + \pi^2 )\frac{s^2+u^2}{2t}
\\
    \delta^1_{(\gamma Z)} &=&
  8 u^2 ( 4 I_{\gamma Z} -  {\hat I}_{\gamma Z} ) \nonumber\\
     \delta^2_{(\gamma Z)} &=&
  8 s^2 ( I_{\gamma Z} -  4 {\hat I}_{\gamma Z} )
\nonumber 
\end{eqnarray}
The logs in the electromagnetic box diagrams are 
\beq
L_s=\ln\frac{s}{|t|},\ L_x=\ln\frac{u}{t}
\eeq
and the scalar integrals in ${\gamma Z}$-parts are
\begin{eqnarray}
I_{\gamma Z} &=& 
\frac{1}{2 \sqrt{-u}} \int_0^1 z dz \int_0^1 dx
\frac{1}{\sqrt{\beta}}
      \ln\left|\frac{ xz\sqrt{-u}+\sqrt{\beta} }{ xz\sqrt{-u}-\sqrt{\beta} }\right|
\nonumber\\
\hat{I}_{\gamma Z} &=& I_{\gamma Z}|_{u \rightarrow -s}
\label{eq:IgammaZ} \\
 \beta &=& -ux^2z^2 +4(1-z)(tz(x-1)+m_Z^2)
\nonumber
\end{eqnarray}
Over a fairly broad kinematic range of interest 
$10^{-4}~\mathrm{GeV}^2 \le |u|,|t| \le 1~\mathrm{GeV}^2$, 
Eq.~(\ref{eq:IgammaZ}) can be integrated
numerically and approximated by the following expression to better
than 1\% precision:
\beq
I_{\gamma Z}\approx \frac{0.123}{m_Z^2}\left(1.64 + 
   \ln\frac{m_Z^2}{-u}\right) + \frac{0.61 t}{m_Z^4}\left(0.102 + 
   \ln\frac{m_Z^2}{-u}\right)\ .
\label{eq:IgammaZapprox}
\eeq

Separating the infrared-divergent and finite virtual contributions,
we can write 
\begin{equation}
\sigma^{V}=
\sigma_{IR}^{V}
+\sigma^{V}(\lambda^2 \rightarrow s ),
\end{equation}
where we introduce the finite photon mass $\lambda$ to regulate the
IR divergence. The IR-divergent part proportional to the Born cross
section is 
\begin{equation}
\sigma_{IR}^{V}=
-\frac{2 \alpha}{\pi} \ln \frac{s}{\lambda^2}
(\ln \frac{tu}{m^2s}-1)\sigma^0\ .
\label{IRV}
\end{equation}
%


\section{Bremsstrahlung contribution}

\begin{figure}
\begin{tabular}{cc}
\epsfig{file=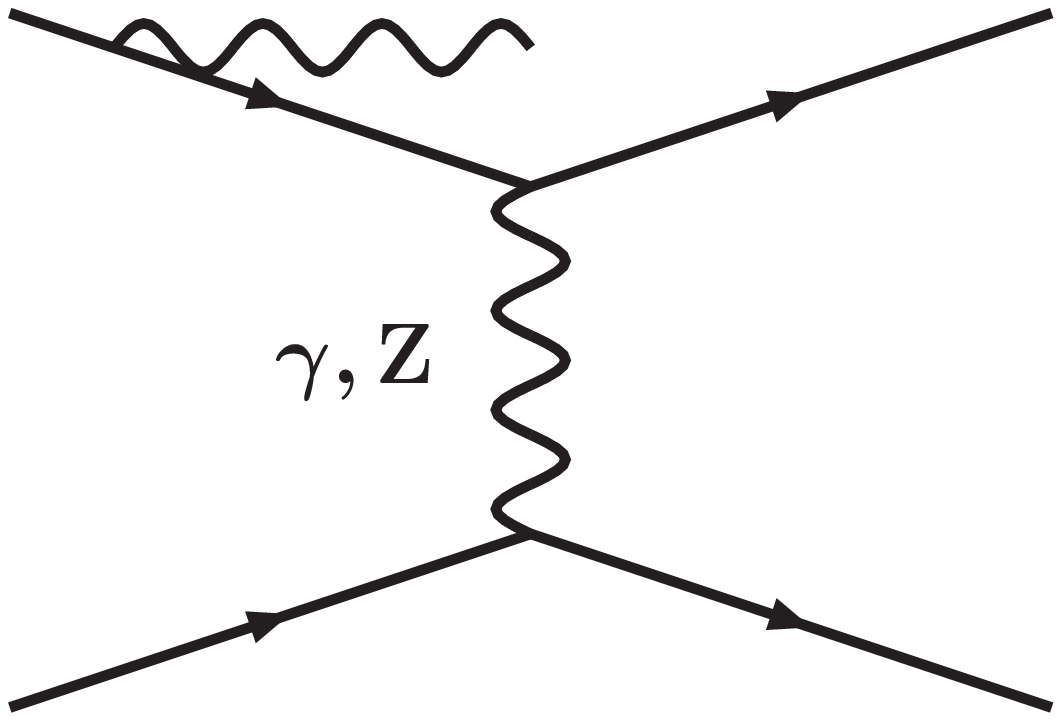,width=1.5in}  &
\epsfig{file=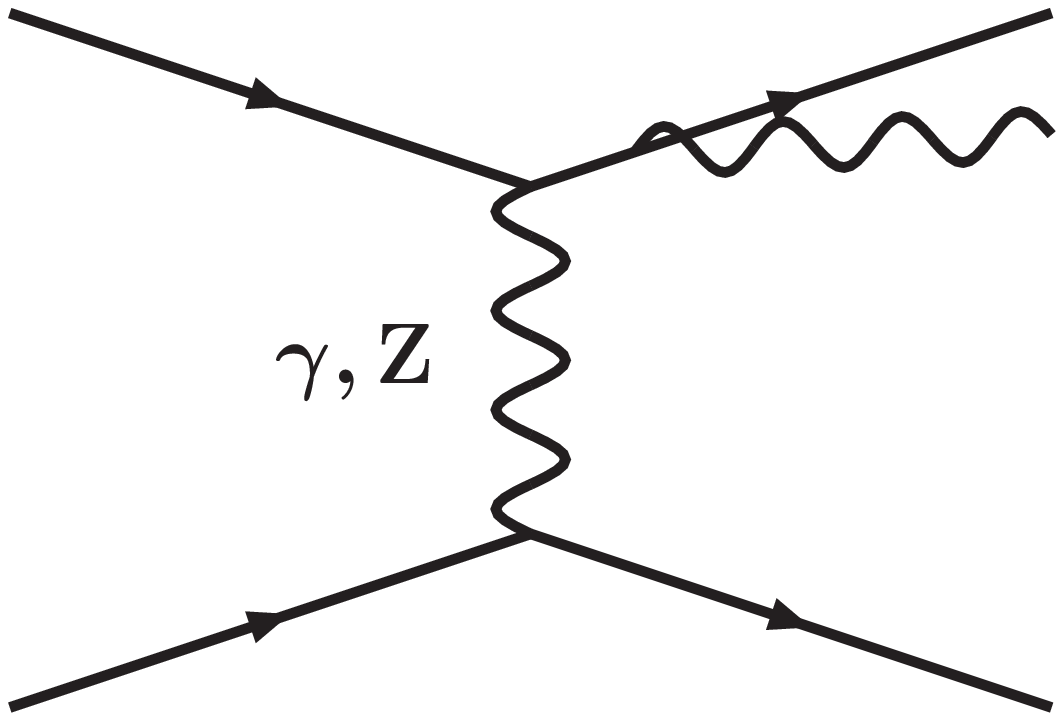,,width=1.5in} \\
\epsfig{file=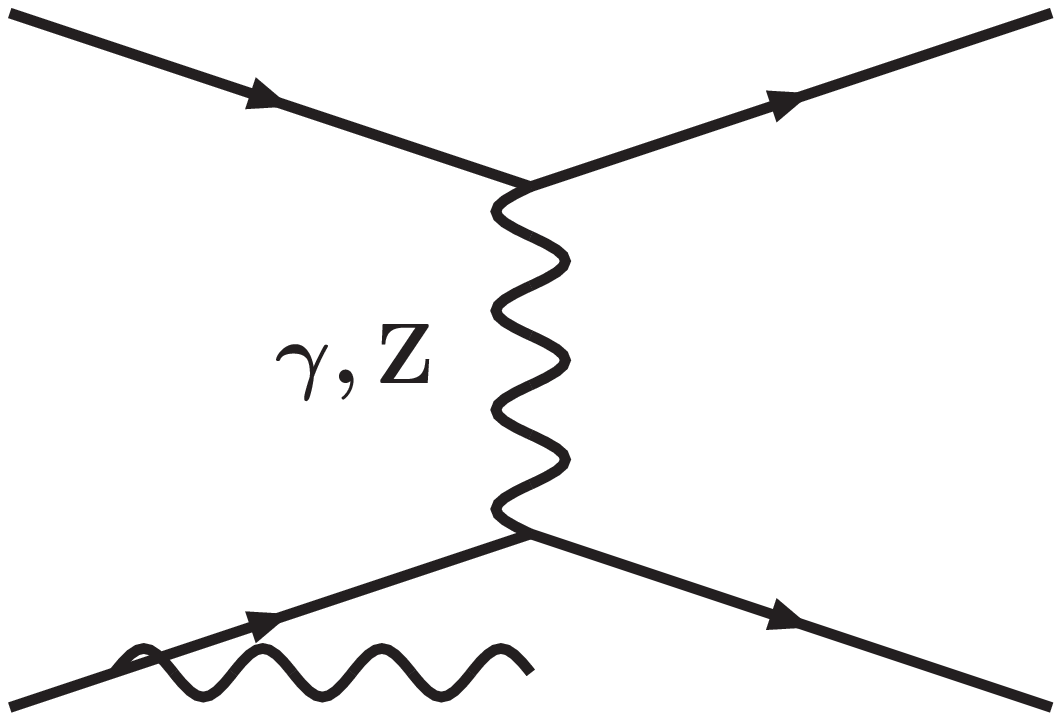,width=1.5in} &
\epsfig{file=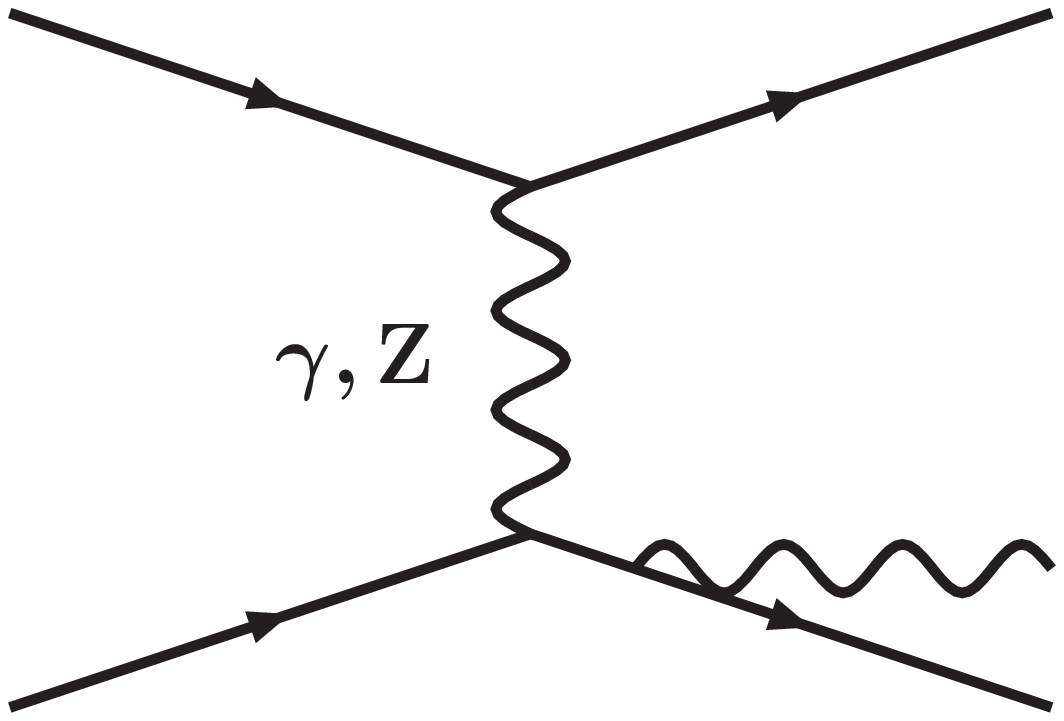,width=1.5in} \\
\end{tabular}
\caption{
$t$-channel bremsstrahlung contributions to process 
$e^-e^- \rightarrow e^-e^- \gamma$. $u$-channel contributions are
  obtained by crossing. 
}
\label{3f}
\end{figure}

The complete leading order radiative corrections need to include the
inelastic processes that correspond to the real photon emission 
$e^{-}e^{-} \to e^{-}e^{-} \gamma$ ($R$-contributions, or real photon
bremsstrahlung). The diagrams are shown in \figref{3f} (plus crossed
terms for a total of 16 diagrams with $\gamma$ and $Z^0$ propagators). 
Let $k$ be the four-momentum of the emitted photon. The differential
cross section is given by 
\bea
\sigma^R
= - \frac{\alpha^3}{4s\pi}
   \int_0^{v^{\max}}dv
 \int\frac{d^3k}{k_0} \delta[(k_1+p_1-k_2-k)^2-m^2] 
 \sum_{j,i=1,4} M_{ij}^R (-1)^{i+j}\ , 
\label{rrr}
\eea
where indices $i$ and $j$ refer to a particular contribution to the cross
section ($u$ and $t$ channels with $Z^0$ or $\gamma$ exchange): 
\begin{equation}
i,j = (1;2;3;4) = (\gamma t;\gamma u;Zt;Zu).
\label{ij}
\end{equation}
We also use a somewhat unconventional set of kinematic variables 
\begin{eqnarray}
z   &=& 2kk_2 \nonumber \\
z_1 &=& 2kk_1 \nonumber\\ 
t_1 &=& (p_2-p_1)^2 \nonumber \\
v   &=& 2kp_2=s+u+t-4m^2 \\
v_1 &=& 2kp_1=s+u+t_1-4m^2 \nonumber \\
z_2 &=& (k_1-p_2)^2=u-v+z_1\ ,\nonumber
\end{eqnarray}
which satisfy the following relations:
\begin{equation}
v_1-v=z-z_1=t_1-t \ .
\end{equation}

The integration variable $v$ in Eq.~(\ref{rrr}) describes the inelasticity
of the reaction, the deviation from the 2-body kinematic constraint 
$s+u+t=4m^2$. The kinematically allowed region for variable $v$ is
given by~\cite{chew-low}  
\beq
v \le v^{\rm lim} = \frac{2(s+t-4m^2)}{1+\sqrt{(1-4m^2/s)(1+4m^2/t)}} \approx s+t
\eeq
The limit $v=v^{\rm lim}$ corresponds to the collinear singularity
$u=0$. However, for E158, which only detects 
scattered electrons with laboratory energy 
$E'\geq E_{\rm cut} \approx 11$~GeV, the integration 
region is further restricted to $u^{\max} =2m(m-E_{\rm cut})$ and
\begin{equation}
v^{\max}=s+t+u^{\max}-4m^2\, .
\label{eq:vmax}
\end{equation}

The squares of the matrix elements $M_{ij}^R$ are given by
\beq
M^R_{ij}=
(M^R_{ij})_{zz}+(M^R_{ij})_{zv}+(M^R_{ij})_{vz}+(M^R_{ij})_{vv}
\eeq
for $ij=13,31,11,33$, and 
\beq
M^R_{ij}=
(M^R_{ij})_{f}+(M^R_{ij})_{l}+(M^R_{ij})_{tu}+(M^R_{ij})_{s}
\eeq
for $ij=12,14,32,34$,
where the traces of the appropriate
$\gamma$-matrix combinations are multiplied by the density matrices and
the corresponding propagators
\begin{widetext}
\begin{eqnarray}
{(M^R_{ij})}_{zz} &=&
 {\mbox Tr}
  [ G_1^{\mu \alpha} \rho^{ij}(k_1) {G_1^{\nu \alpha}}^T \Lambda(k_2) ]
  {\mbox Tr}
  [ \gamma_{\mu} \rho^{ij}(p_1) \gamma_{\nu} \Lambda(p_2) ]
  D^{it_1}D^{jt_1},
\nonumber \\
{(M^R_{ij})}_{zv}&=&
 {\mbox Tr}
  [ G_1^{\mu \alpha} \rho^{ij}(k_1) \gamma_{\nu} \Lambda(k_2) ]
  {\mbox Tr}
  [ \gamma_{\mu} \rho^{ij}(p_1) {G_2^{\nu \alpha}}^T \Lambda(p_2) ]
  D^{it_1}D^{jt},
\nonumber \\
{(M^R_{ij})}_{vz}&=&
 {\mbox Tr}
  [ G_2^{\mu \alpha} \rho^{ij}(p_1) \gamma_{\nu} \Lambda(p_2) ]
  {\mbox Tr}
  [ \gamma_{\mu} \rho^{ij}(k_1) {G_1^{\nu \alpha}}^T \Lambda(k_2) ]
  D^{it}D^{jt_1},
\nonumber \\
{(M^R_{ij})}_{vv}&=&
 {\mbox Tr}
  [ G_2^{\mu \alpha} \rho^{ij}(p_1) {G_2^{\nu \alpha}}^T \Lambda(p_2) ]
  {\mbox Tr}
  [ \gamma_{\mu} \rho^{ij}(k_1) \gamma_{\nu} \Lambda(k_2) ]
  D^{it}D^{jt},
\nonumber \\
{(M^R_{ij})}_{f}&=&
 {\mbox Tr}
  [ G_1^{\mu \alpha} \rho^{ij}(k_1) {G_3^{\nu \alpha}} \Lambda(p_2)
  \gamma_{\mu} \rho^{ij}(p_1) \gamma_{\nu} \Lambda(k_2) ]
  D^{it_1}D^{ju},
\nonumber \\
{(M^R_{ij})}_{l}&=&
 {\mbox Tr}
  [ G_1^{\mu \alpha} \rho^{ij}(k_1) \gamma_{\nu} \Lambda(p_2)
  \gamma_{\mu} \rho^{ij}(p_1) {G_4^{\nu \alpha}} \Lambda(k_2) ]
  D^{it_1}D^{jz_2},
\nonumber \\
{(M^R_{ij})}_{tu}&=&
 {\mbox Tr}
  [ \gamma_{\mu} \rho^{ij}(k_1) G_3^{\nu \alpha} \Lambda(p_2)
    G_2^{\mu \alpha} \rho^{ij}(p_1) \gamma_{\nu} \Lambda(k_2) ]
  D^{it}D^{ju},
\nonumber \\
{(M^R_{ij})}_{s}&=&
 {\mbox Tr}
  [ \gamma_{\mu} \rho^{ij}(k_1) \gamma_{\nu} \Lambda(p_2)
    G_2^{\mu \alpha} \rho^{ij}(p_1) G_4^{\nu \alpha} \Lambda(k_2) ]
  D^{it}D^{jz_2},
\label{term-r}
\end{eqnarray}
\end{widetext}
where
\begin{equation}
\Lambda(p)=\hat p+m, \ \hat p=\gamma^{\mu}p_{\mu},
\end{equation}
\begin{equation}
 G_1^{\mu \alpha} =
   \gamma^{\mu}\frac{2k_1^{\alpha}-\hat k \gamma^{\alpha}}{-z_1} +
   \frac{2k_2^{\alpha}+\gamma^{\alpha}\hat k}{z}\gamma^{\mu},
\end{equation}
\begin{equation}
 G_2^{\mu \alpha} =
   \gamma^{\mu}\frac{2p_1^{\alpha}-\hat k \gamma^{\alpha}}{-v_1} +
   \frac{2p_2^{\alpha}+\gamma^{\alpha}\hat k}{v}\gamma^{\mu},
\end{equation}
\begin{equation}
 G_3^{\nu \alpha} =
 \frac{2k_1^{\alpha}-\gamma^{\alpha}\hat k}{-z_1}\gamma^{\nu} +
 \gamma^{\nu} \frac{2p_2^{\alpha}+\hat k \gamma^{\alpha}}{v},
\end{equation}
\begin{equation}
 G_4^{\nu \alpha} =
 \frac{2p_1^{\alpha}-\gamma^{\alpha}\hat k}{-v_1}\gamma^{\nu} +
 \gamma^{\nu} \frac{2k_2^{\alpha}+\hat k \gamma^{\alpha}}{z}.
\end{equation}

Expressions for $M^R_{ij}$ for other values of $i$ and $j$
(\eg\ $\{ij\} = 22,44,24,42$, etc.) can be obtained from the symmetry of
expressions in Eq.~(\ref{term-r}):
\begin{eqnarray}
{(M^R_{24})}_{zz} &=&
 \left. {(M^R_{13})}_{zz} \right|_{k_2 \leftrightarrow p_2} =
 \left. {(M^R_{13})}_{zz} 
         \right|_{k_2 \leftrightarrow p_2,k_1 \leftrightarrow p_1} 
\nonumber \\
&=&  \left. {(M^R_{13})}_{vv} \right|_{k_1 \leftrightarrow p_1} =
     \left. {(M^R_{13})}_{vv} \right|_{t \leftrightarrow u},
\end{eqnarray}
and
\begin{eqnarray}
{(M^R_{24})}_{vv} &=&
     \left. {(M^R_{13})}_{zz} \right|_{t \leftrightarrow u}.
\end{eqnarray}
Cases $\{ij\} = 22,44$ are analyzed in a similar manner. The symmetry noted
above is less apparent in the interference terms (indices $zv$ and
$vz$):
\begin{eqnarray}
{(M^R_{24})}_{zv}+{(M^R_{42})}_{vz} &=&
     \left. \Bigl[ {(M^R_{13})}_{vz} + {(M^R_{31})}_{zv} \Bigr]
     \right|_{t \leftrightarrow u}
\end{eqnarray}
and likewise for other parts in Eq.~(\ref{rrr}).

\subsection{Infrared Divergences in Bremsstrahlung Contributions}

Next we need to address the issue of the infrared divergences in the
bremsstrahlung cross section. According to the prescription of Bardin and
Shumeiko~\cite{covar}, we find the infrared-divergent parts in the
squares of the matrix elements that are proportional to the
corresponding Born contributions:
\begin{eqnarray}
{(M^R_{ij})}_{zz}^{IR} &=&
4(\frac{m^2}{z^2}+\frac{m^2}{z_1^2}+\frac{t}{zz_1})M_{ij},
\label{mirr} \\
{(M^R_{ij})}_{zv}^{IR} +
{(M^R_{ij})}_{vz}^{IR} &=&
4(\frac{u}{zv_1}+\frac{s}{zv}+\frac{s}{z_1v_1}+\frac{u}{z_1v})M_{ij},
\nonumber \\
{(M^R_{ij})}_{vv}^{IR} &=&
4(\frac{m^2}{v_1^2}+\frac{m^2}{v^2}+\frac{t}{v_1v})M_{ij},
\nonumber \\
{(M^R_{ij})}_{f}^{IR} &=&
(\frac{4m^2}{z_1^2}+\frac{2u}{z_1v}+\frac{2t}{zz_1}+\frac{2s}{zv})M_{ij},
\nonumber \\
{(M^R_{ij})}_{l}^{IR} &=&
(\frac{2s}{z_1v_1}+\frac{2t}{z_1z}+\frac{2u}{zv_1}+\frac{4m^2}{z^2})M_{ij},
\nonumber \\
{(M^R_{ij})}_{tu}^{IR} &=&
(\frac{2s}{z_1v_1}+\frac{2u}{z_1v}+\frac{2t}{vv_1}+\frac{4m^2}{v^2})M_{ij},
\nonumber \\
{(M^R_{ij})}_{s}^{IR} &=&
(\frac{4m^2}{v_1^2}+\frac{2u}{v_1z}+\frac{2t}{vv_1}+\frac{2s}{vz})M_{ij},
\nonumber
\end{eqnarray}
where
\begin{eqnarray}
M_{11} &=& D^{\gamma t} D^{\gamma t}M_{e}^{\gamma \gamma \gamma \gamma}
\nonumber\\
M_{13} &=& D^{\gamma t} D^{Z t}M_{e}^{\gamma Z \gamma Z}
\nonumber \\
M_{31} &=& D^{\gamma t} D^{Z t}M_{e}^{Z\gamma Z \gamma} \nonumber\\
M_{33} &=& D^{Z t} D^{Z t}M_{e}^{ZZZZ} \nonumber \\
M_{12} &=& D^{\gamma t} D^{\gamma u}M_{o}^{\gamma \gamma \gamma \gamma}
\nonumber\\
M_{14} &=& D^{\gamma t} D^{Z u}M_{o}^{\gamma Z \gamma Z} \nonumber \\
M_{32} &=& D^{\gamma u} D^{Z t}M_{o}^{Z\gamma Z \gamma} \nonumber\\
M_{34} &=& D^{Z t} D^{Z u}M_{o}^{ZZZZ} \ .
\end{eqnarray}

The complete $R$-contribution to the cross section is
infrared-divergent, but can be separated into an infrared-infinite
part $\sigma_{IR}^{R}$ and the finite part
$\sigma^R_{F}$:
\begin{equation}
\sigma^R=  \sigma^R_{IR} + \sigma^R_{F}\ .
\label{sir}
\end{equation}
The IR-divergent part of the bremsstrahlung matrix elements,
proportional to the Born contributions, can be constructed from
Eq.~(\ref{mirr}) according to Eq.~(\ref{rrr}). The finite contribution
to the cross section is then obtained by subtraction
\begin{equation}
{(M^R_{ij})}^{F} = {M^R_{ij}} - {(M^R_{ij})}^{IR},
\label{vych}
\end{equation}

The infrared-divergent part of Eq.~(\ref{sir}), integrated over
variables $k$ and $v$ is given in terms of a finite photon mass
$\lambda$ by~\cite{my3}
\begin{equation}
\sigma_{IR}^{R}=
\frac{2 \alpha}{\pi} \ln \frac{(v^{\max})^2}{s\lambda^2}
(\ln \frac{tu}{m^2s}-1)\sigma^0\ .
\label{IRR}
\end{equation}

The integration over the phase space of the bremsstrahlung photon is
performed analytically, and integration over variable $v$ is done
numerically due to complexity of the integral expressions. The photon
phase space integral can be written as~\cite{bu-ka}
\begin{eqnarray}
\label{int}
I[A]&=&
 \frac{1}{\pi}\int\frac{d^3k}{k_0} \delta[(k_1+p_1-k_2-k)^2-m^2][A]
=
\frac{1}{\pi} \int \limits _{t_1^{\min}}^{t_1^{\max}}\ dt_1
     \int \limits _{{z}^{\min}}^{{z}^{\max}}
     \frac{dz}{\sqrt{R_z}}[A],
\end{eqnarray}
where $R_z$ is the Gramm determinant (modulo $-1$), and can be
parameterized as a second-order polynomial in $z$ 
as 
\begin{equation}
 R_z = A_z z^2 + 2 B_z z + C_z\ .
\end{equation}
Coefficients $A_z$, $B_z$, and $C_z$ are given by the following
expressions:
 \begin{eqnarray}
 A_z &=& 4m^2t-(v-t)^2, \nonumber \\
 B_z &=& Et_1+F \nonumber \\
 E   &=& v(u-2m^2)-st,\ F=t(2m^2v+st-sv),
\nonumber \\
 C_z &=& -(At_1^2+2Bt_1+C), \\
 A &=&(s-v)^2-4m^2s \ , \nonumber \\
 B &=& -st(s-v-4m^2)-2m^2v^2\ ,\nonumber\\
 C &=& st^2(s-4m^2)\ .\nonumber
 \end{eqnarray}
Integration limits $z^{\min/\max}$ and $ t_1^{\min/\max}$ 
are solutions of the equations 
$R_z=0$ and $z^{\min}=z^{\max}$:
\bea
 z^{\min/\max}&=&\frac{-{B_z} \pm \sqrt{B_z^2-A_zC_z}}{A_z}
\\
  t_1^{\min/\max}&=&\frac{v(t-v)+2m^2t \mp v\sqrt{-A_z}}{2(v+m^2)}
\eea
This set of variables makes integration more convenient.

Expressions for ${(M^R_{ij})}^{F}$ are computed using symbolic
manipulation program {\tt REDUCE\/}~\cite{reduce} but are excessively 
complex to be listed here. They 
are available as subroutines in the Fortran program 
{\tt rcAPV\/}\footnote{Fortran program {\tt rcAPV\/} is available from
  the authors upon request.}.  
The relevant integrals, listed in the Appendix
of Ref.~\cite{zyk}, were computed both analytically and numerically. 

\section{Cancellation of Infrared Singularities}

Adding infrared-divergent parts of $V$- and $R$-contributions given in
Eq.~(\ref{IRV}) and Eq.~(\ref{IRR})) together with the IR-finite
pieces, we obtain the finite expression for the radiatively-corrected
cross section, free of non-physical parameters:
\begin{eqnarray}
\sigma &=& \sigma^0 + \sigma^C
\nonumber \\
\sigma^C&=& \sigma_{{IR}}^{V} +\sigma_{{IR}}^R +\sigma^R_F 
         + \sigma^{Ver}_F + \sigma^B_F
\label{can}\\
&=&
\frac{\alpha}{\pi}
( 4 \ln \frac{v^{{max}}}{m\sqrt{s}}
  (\ln \frac{tu}{m^2s}-1) + \delta_1^S + \delta_1^H
) \sigma^0 
         + \sigma^{Ver}_F + \sigma^B_F + \sigma^R_{F}\ ,
\nonumber
\end{eqnarray}
where
\begin{widetext}
\begin{eqnarray}
      \delta_1^S &=& \ln\frac{s(s+t)}{m^4}
                  - \frac{1}{2}l_m \ln\frac{s^2(s+t)^2}{-tm^6}
                  - \frac{1}{2}l_r^2 - 2l_rl_m+l_m-l_m^2-\frac{\pi^3}{3}+1,
\nonumber \\
      \delta_1^H  & = & \int\limits_0^{v^{max}} dv
  \Bigl( -\frac{2}{v}\ln(1-\frac{v}{s})
    +\frac{2}{v}\ln(1-\frac{v}{t})
    +\frac{2}{v}\ln(1-\frac{v}{s+t})
    -\frac{1}{v}\ln(1+\frac{v}{m^2})
\nonumber \\
    && +\frac{2}{s+t-v}\ln\frac{s+t-v}{m^2}
    -\frac{1}{s-v}\ln\frac{(s-v)^2}{m^2\tau}
    -\frac{1}{v-t}\ln\frac{(v-t)^2}{m^2\tau}
    -\frac{1}{\tau} \Bigr),
\nonumber \\
&& l_m=\ln\frac{-t}{m^2},\ l_r=\ln\frac{s+t}{s},\ \tau\equiv v+m^2,
\end{eqnarray}
\end{widetext}

\section{Results and Discussion}
\label{sec:results}

\subsection{Numerical Results}

In the following, we evaluate the effect of the bremsstrahlung
radiative corrections on the parity-violating asymmetry \ALR\ in
the scattering of the longitudinally polarized electrons off
unpolarized target electrons. We consider the kinematic conditions that
correspond to the experimental setup of the SLAC E158
experiment~\cite{E158}, \ie\ beam energies of $45$ and $48$ GeV. 
E158 
setup is such that the radiated photon is not detected. Moreover, the
scattered electrons are only detected if their energy is above the
threshold $E' \geq 11$~GeV; this restriction limits the range of
integration over variable $v$ as given in Eq.~(\ref{eq:vmax}). 

The relative corrections to the cross section and asymmetry can be defined as
\begin{equation}
\delta\sigma = \frac{\sigma-\sigma^0}{\sigma^0}\ ,\  
\delta A=\frac{\ALR-\ALR^0}{\ALR^0},
\label{dA}
\end{equation}
where $A_0$ is the Born asymmetry, and $A^{RC}$ is the radiatively
corrected asymmetry. 

\begin{figure}
\begin{center}
\epsfig{file=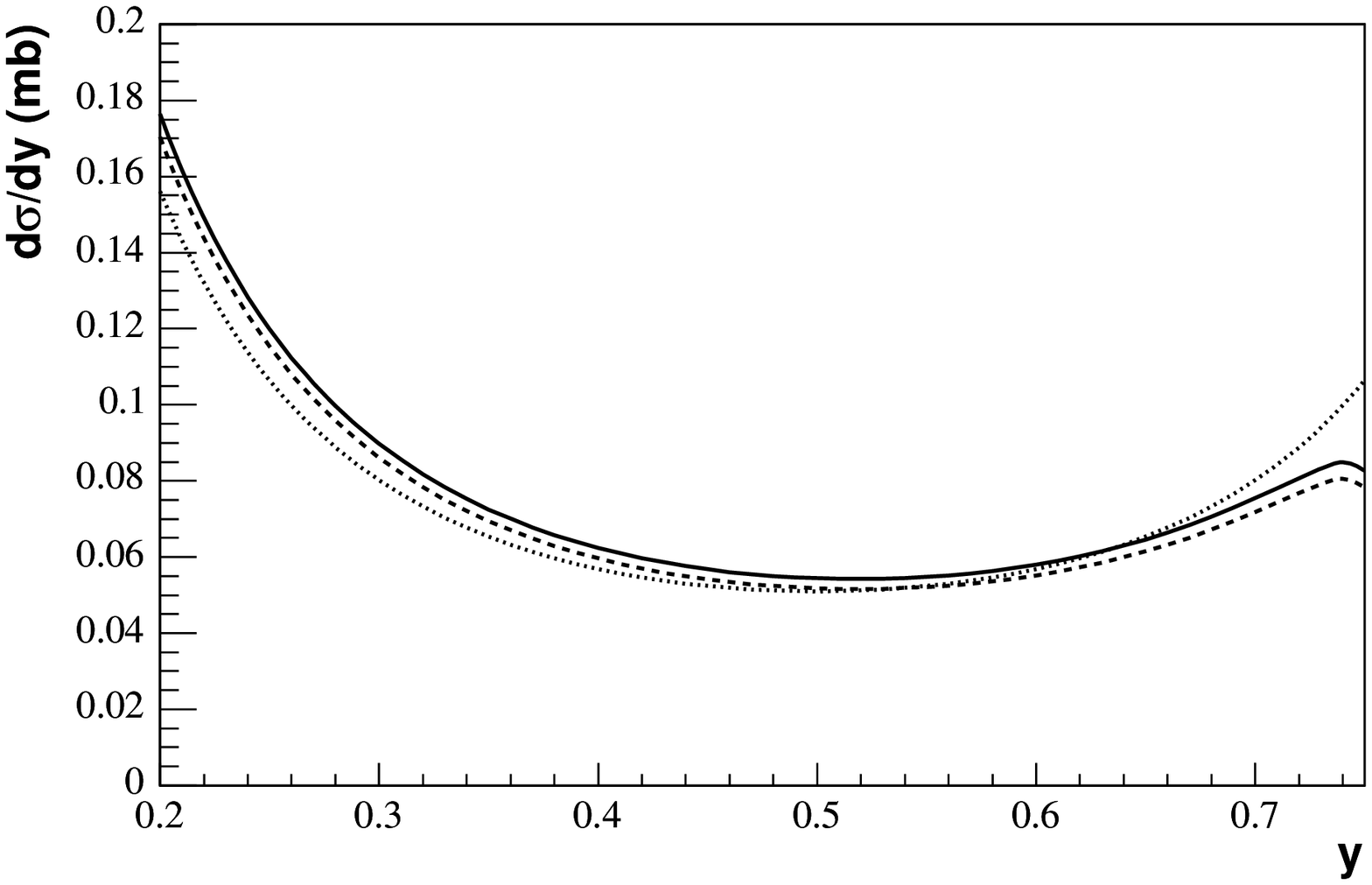,width=3.5in} 
\epsfig{file=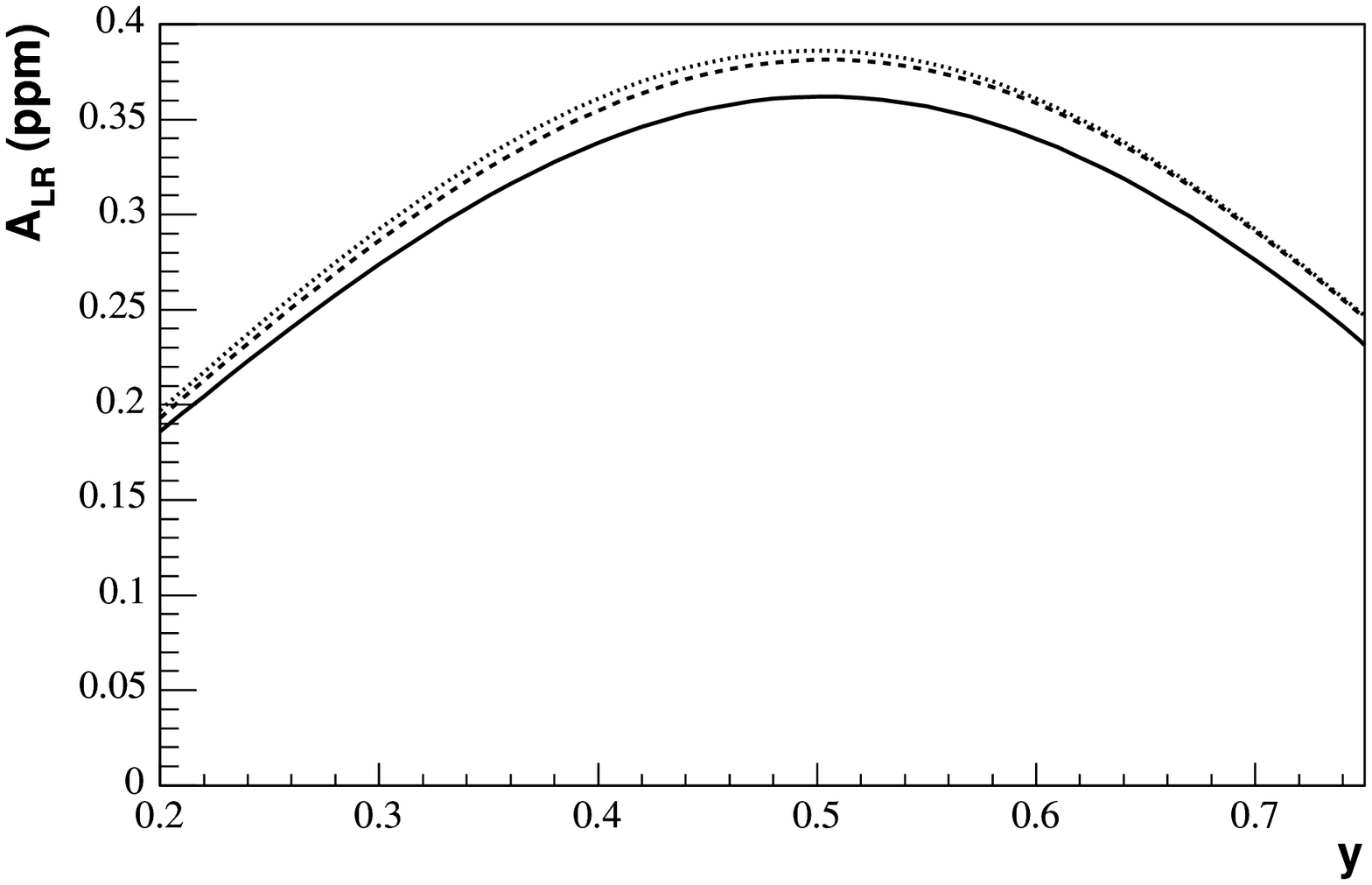,width=3.5in} 
\caption{
Differential cross section (left) and asymmetry \ALR\ (right) as a
function of $y$ for beam energy $E=45$ GeV.  
Dotted line shows the Born cross section and asymmetry, 
effects of bremsstrahlung corrections are displayed by dashed lines, 
and the solid line is the result of the
full QED corrections. The plots are cut off at
$y\approx 0.75$ by the E158 spectrometer acceptance ($E'>11$ GeV).
}
\label{fig:sigAsymVsY}
\end{center}
\end{figure}

\begin{figure}
\begin{center}
\epsfig{file=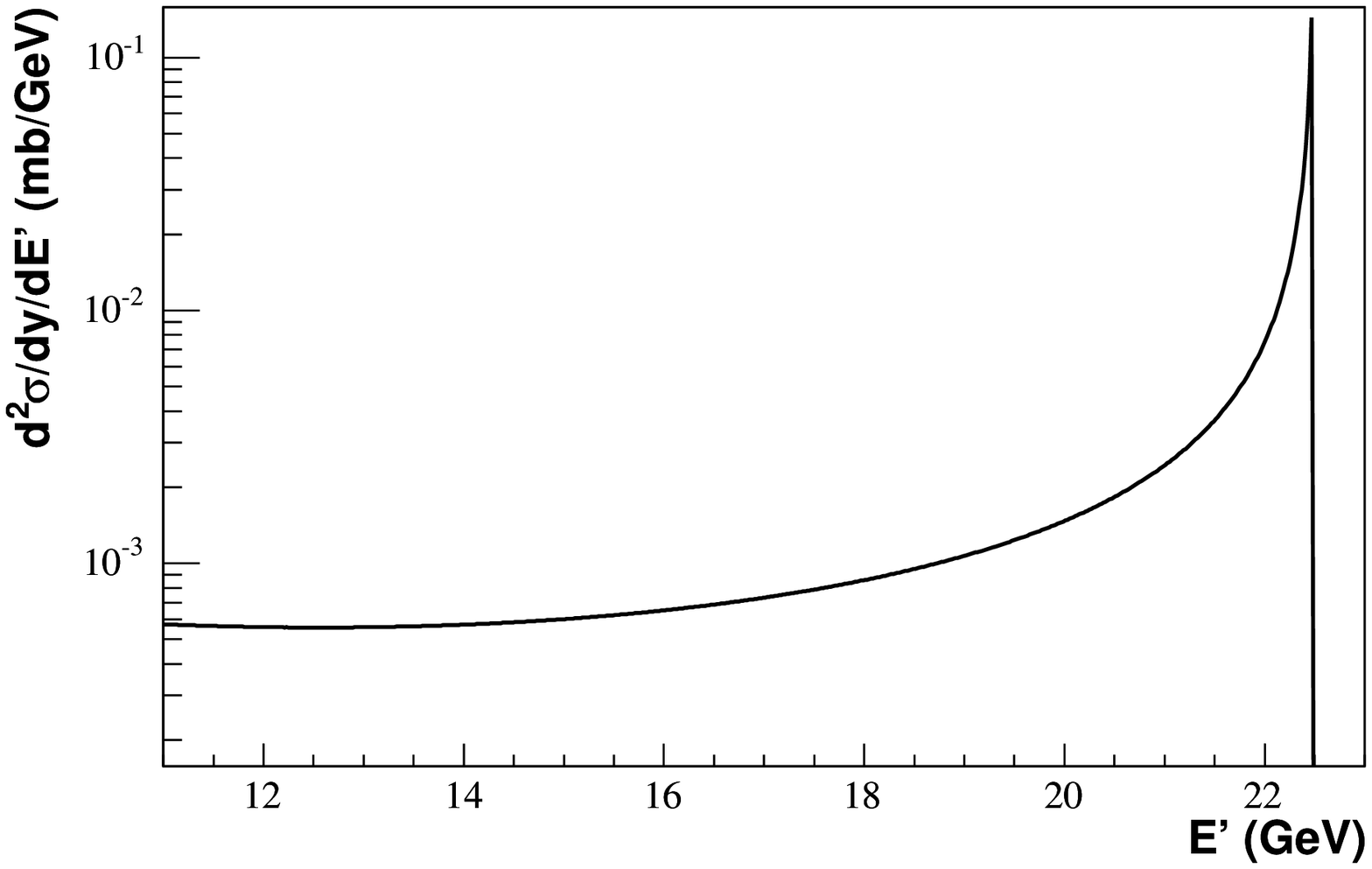,width=3.5in} 
\epsfig{file=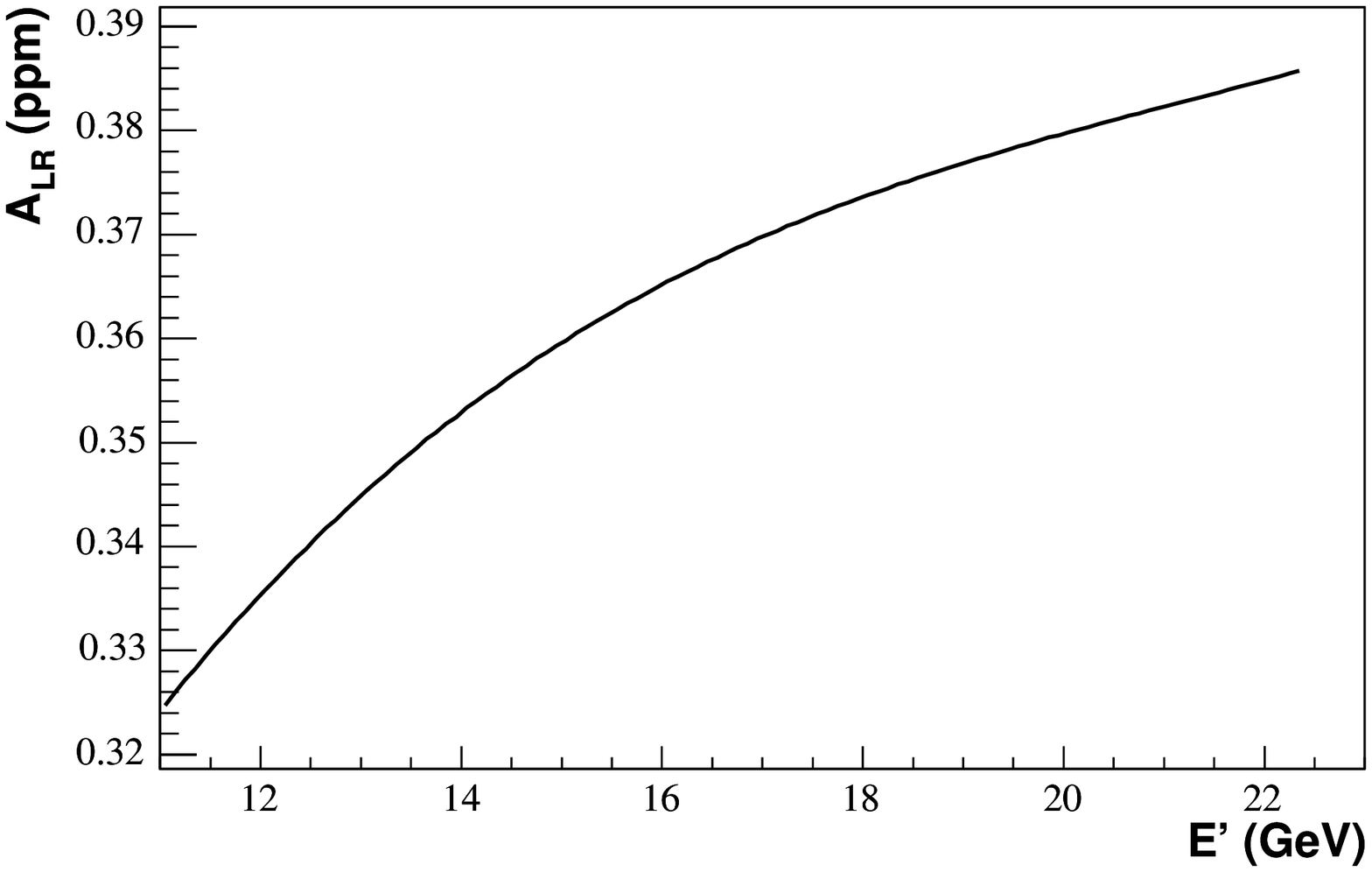,width=3.5in} 
\caption{
\Moller\ scattering cross section (left) and asymmetry \ALR\ (right)
as a function of scattered electron energy $E^{'}$ for 
beam energy $E=45$ GeV and $y=0.5$. 
}
\label{fig:sigAsymVsE}
\end{center}
\end{figure}

\begin{figure}
\begin{center}
\epsfig{file=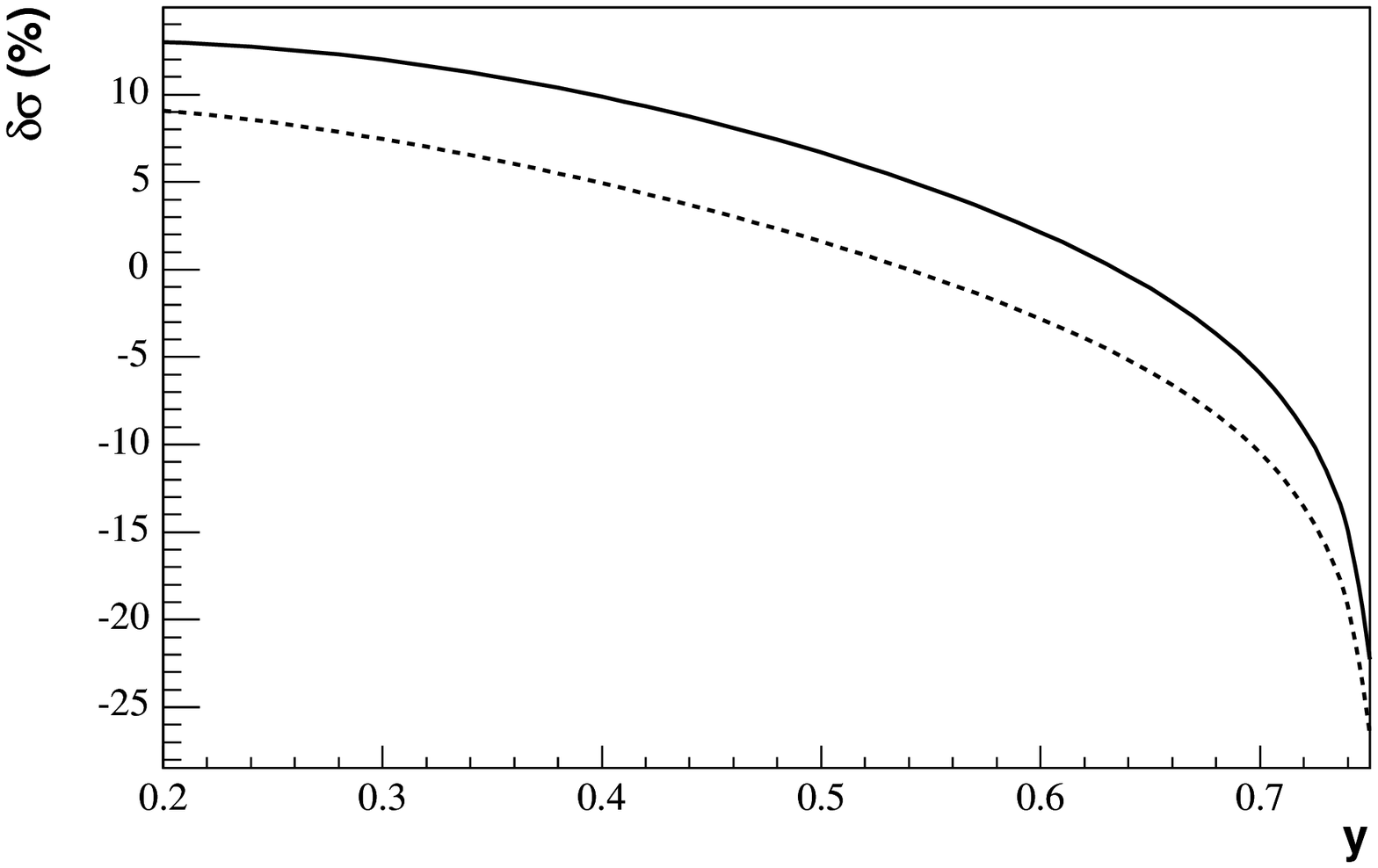,width=3.5in} 
\epsfig{file=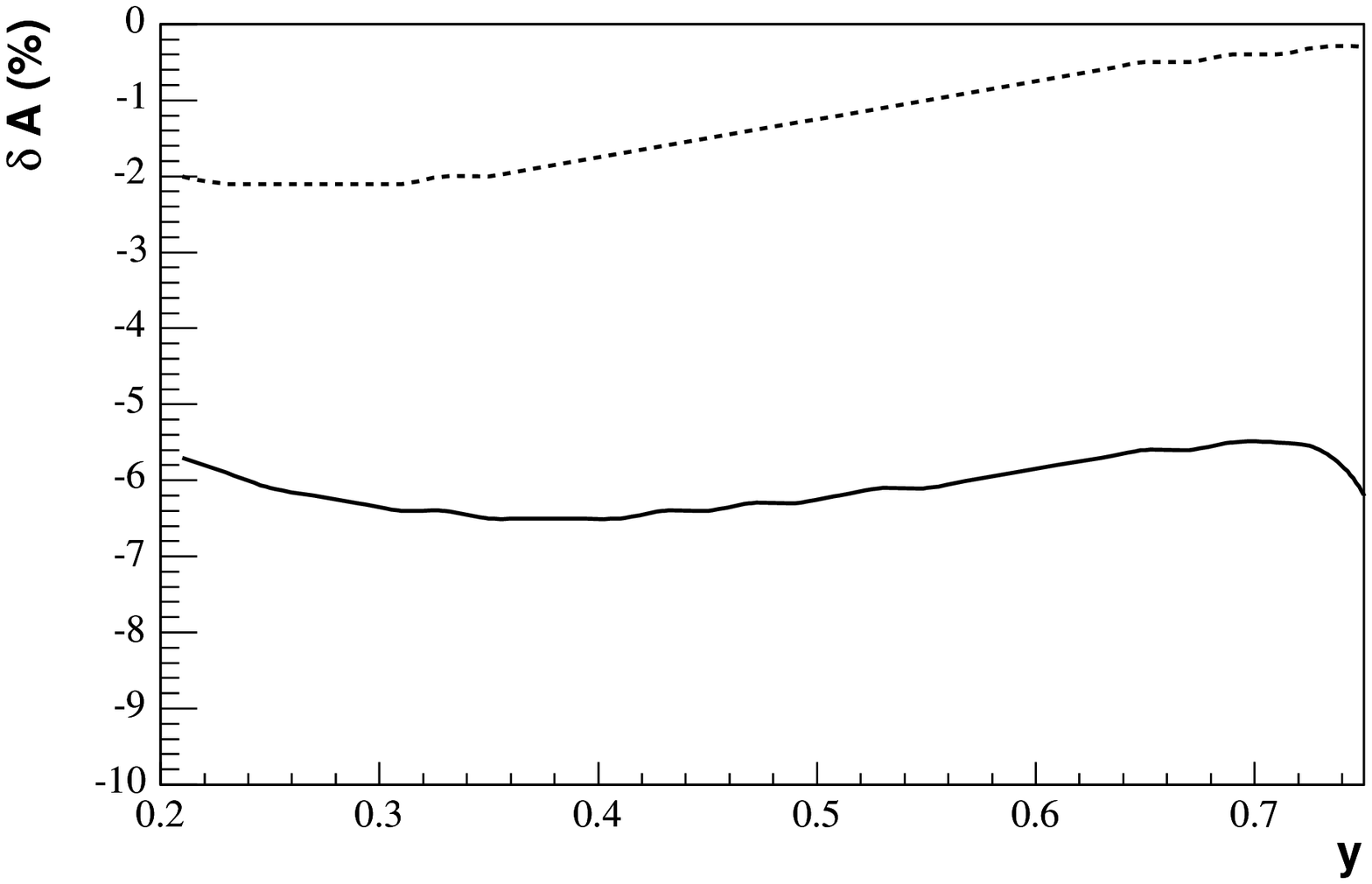,width=3.5in} 
\caption{
Relative corrections to the \Moller\ scattering  cross section (left)
and to the parity-violating asymmetry \ALR\ (right) as a
function of $y$. Dashed line includes
bremsstrahlung corrections, and the solid line is the result of the
full first-order QED calculation.
}
\label{fig:corr}
\end{center}
\end{figure}

The Born cross section $\sigma^0$, the radiatively corrected cross
sections $\sigma^\mathrm{brem}$ (which includes only soft and hard
bremsstrahlung corrections and results of IR cancellation) and
$\sigma^{QED}$ (full QED corrections), as well as the asymmetry \ALR\
are shown as a function of 
variable $y$ in \figref{fig:sigAsymVsY} for beam energy of
$E=45$~GeV. 
\figref{fig:sigAsymVsE} shows the double-differential cross section
$d^2\sigma/dE'/dy$ and the asymmetry $A_{LR}$ as a function of the
scattered electron energy in the lab frame $E'$ for a fixed
$y=0.5$. This double-differential cross section is used to properly
average the radiative corrections over the experimental acceptance. 
The corrections to cross section and
asymmetry are shown in Fig.~\ref{fig:corr}.
The numerical precision of the corrections is about $0.1\%$. 

We find that at fixed value of $y=0.5$ and $E=45$~GeV
($Q^2=0.023~\mathrm{GeV}^2$), hard bremsstrahlung reduces the value of
the parity-violating asymmetry by $\approx 1\%$. Contribution from the
$\gamma\gamma$ and $\gamma Z$ box diagrams is also negative and
reduces the asymmetry by $\approx 5\%$, so that the total QED
correction at $y=0.5$ and $E=45$~GeV is $-6.2\%$. 


At fixed scattering angles,
radiative effects move events towards lower values of $y$. Therefore,
even though at a fixed value of $y=0.5$ the hard bremsstrahlung
correction is negative, the net change of the asymmetry, integrated
over E158 acceptance $-0.5\le\cos\Theta \le 0$ at $E=45$~GeV is 
$\langle \delta A^\mathrm{brem} \rangle \approx +1\%$. The full QED
correction, including $\gamma\gamma$ and $\gamma Z$ boxes, is 
$\langle \delta A^{QED}\rangle = -4.5\%$. 

\subsection{Factorization of QED Radiative Corrections and NLO Uncertainties}

At leading order, corrections to the parity-violating asymmetry
from the diagrams involving photons (\ie\ soft and hard
bremsstrahlung, $\gamma\gamma$ and $\gamma Z$ boxes, and photonic
vertex diagrams) are proportional to the product $v^Z a^Z$. In the
OS scheme, these contributions are strictly proportional to the
Born asymmetry. In other words, in the OS scheme, the relative
corrections to the asymmetry due to QED diagrams are independent of
the weak mixing angle, and can be factorized out. We can write
\beq
\ALR^{QED} = \mathcal{A}^0(Q^2,y) (1+\delta A^{QED}) \left(1-4\sin^2\theta_W\right)\, 
\label{eq:ALRqed}
\eeq
where $\mathcal{A}^0(Q^2,y)$ is the Born analyzing power defined in
Eq.~(\ref{eq:ALRtree}) and $\delta A^{QED}$ is the relative radiative
correction. For $E=45$~GeV and $y=0.5$, full QED correction is 
$\delta A^{QED}=-0.061$, and the average over E158 kinematics (beam energies
of 45 and 48 GeV and $-0.5\le\cos\Theta \le 0$) is 
$\langle \delta A^{QED}\rangle_\mathrm{E158} = -0.043$. 

Virtual corrections, such as vacuum polarization and box and vertex
diagrams with heavy gauge bosons, are not in general proportional to
$v^Z a^Z$, and spoil the simple factorization of
Eq.~(\ref{eq:ALRqed}). Nevertheless, one can 
relate the leading-order asymmetry to the Born-level formula through the set of
multiplicative and additive corrections:
\begin{widetext}
\beq
\ALR(Q^2,y) = \mathcal{A}^0(Q^2,y)\rho(Q^2) (1+\delta A(Q^2,y)) \left(1-4\sin^2\theta_W^{\rm eff}(Q^2)+\Delta(Q^2)\right)\, .
\label{eq:ALRcorr}
\eeq
\end{widetext}
Here the effective mixing angle relevant for a scattering process at
momentum transfer $Q$ is defined through the form-factor $\kappa(Q^2)$
(in OS scheme) or $\hat{\kappa}(Q^2,\mu^2)$ (in \MSbar\ scheme):
\beq
\sinTW^{\rm eff}(Q^2)\equiv \kappa(Q^2)\sinTW^{OS} = 
\hat{\kappa}(Q^2,\mu^2)\sin^2\hat{\theta}_W(\mu^2)\ .
\label{eq:sinTWeff}
\eeq
In \MSbar\ scheme, one typically chooses $\mu=m_Z$. 

All corrections in Eq.~(\ref{eq:ALRcorr}) are of order
$\mathcal{O}(\alpha)$ but have a different physical meaning. 
The
form-factor $\kappa=1+\mathcal{O}(\alpha)$ defines the momentum
dependence (running) of the effective weak mixing angle. We would like
to define it in a process-independent way, such that various 
experimental measurements could be directly compared in terms of
$\sinTW^{\rm eff}$. Typically, $\kappa$ includes contributions from
$\gamma-Z$ mixing and anapole moment diagrams
(\figref{fig:2f}a-c)~\cite{cz-marc}, but may include other
terms~\cite{sirlin,erler-musolf}.  
In definition of Ref.~\cite{cz-marc}, carried out in \MSbar\
scheme with $\sin^2\hat{\theta}_W(m_Z^2) \equiv\hat{s}^2 = 0.23120\pm
0.00015$~\cite{ref:PDG2004},  
$\hat{\kappa}(Q^2=0,\mu^2=m_Z^2) = 1.0298\pm 0.0026$

The form-factor $\rho(Q^2)=1+\mathcal{O}(\alpha)$ is a low-energy
ratio of the neutral weak coupling to the charge coupling. It depends
on the choice of the Fermi constant $G_F$ in Eq.~(\ref{eq:ALRtree}); for
$G_F$ derived from the muon-decay constant $G_\mu$, this correction
is~\cite{cz-marc}  
$\rho = 1.0012\pm 0.0005$. $\rho$ contains logarithmic dependence on
the Higgs mass (we use $m_H = 113_{-40}^{+56}$~GeV~\cite{ref:PDG2004}) and
linear dependence on 
the top quark mass ($m_t = 177\pm 4$~GeV~\cite{ref:PDG2004}). 

The remaining first-order corrections are included in
Eq.~(\ref{eq:ALRcorr}) as factors $\delta A(Q^2,y)$ and $\Delta(Q^2)$.
$\Delta(Q^2)$ typically includes box diagrams with two heavy bosons,
and $\delta A(Q^2,y)$ contains the kinematics-dependent factorizable
QED corrections computed in Section~\ref{sec:results}. 

The only remaining question is evaluating the
next-to-leading order correction uncertainties. Normally, NLO
corrections would be 
of order $\mathcal{O}(\alpha)$ of the LO terms. However, we have to
pay special attention to the logarithmically-enhanced contributions in
the $\gamma Z$ box diagrams, \eg\ terms proportional to $\ln(m_Z^2/s)$
in Eq.~(\ref{eq:IgammaZapprox}). 

The effective $Z^0$-electron coupling, $Q_w = -\rho(1-4\sinTW)$ changes
by about $40\%$ between zero momentum transfer and $Z^0$
pole~\cite{cz-marc}. Since the 
box diagrams in \figref{fig:2f} involve integration over internal
momenta of the photon and $Z^0$ propagators, a complete calculation of
the box diagrams has to take into account the momentum dependence of
$Q_w$ and $\alpha$. 
Strictly speaking, the momentum dependence of the weak or
electromagnetic charges in the box integrals is a next-to-leading
order effect, and is beyond the scope of this work. However, a
judicious choice of the coupling at leading order could reduce the NLO
corrections. 

Czarnecki and Marciano~\cite{cz-marc} have argued that the NLO errors
are reduced if the $\gamma Z$ box diagrams are evaluated in \MSbar\
scheme with the average value of the weak mixing angle 
\beq
\langle \sinTW\rangle \equiv \frac{\sinTW(0)+\sinTW(m_Z^2)}{2}
\eeq
since the relevant integrals over the internal momentum $k$ are dominated
by the poles near $k^2 \sim 0$ and $k^2 \sim m_Z^2$. Similar arguments
apply to the value of the fine structure constant in the $\gamma Z$
box diagrams. Thus, 
following the spirit of this argument, we move the leading
logarithmic contribution to the $\gamma Z$ box diagrams~\cite{cz-marc}
\beq
\Delta_{\gamma Z}(Q^2) =
-\frac{22}{3}\frac{\alpha(Q^2)}{4\pi}\left(1-4\sinTW(Q^2)\right)\ln\frac{m_Z^2}{s}
\eeq
from the multiplicative correction  $\delta A(Q^2,y)$ to the
additive correction $\Delta(Q^2)$. Moreover, we use the average value
$\langle \Delta_{\gamma Z}\rangle\equiv 
(\Delta_{\gamma Z}(0)+\Delta_{\gamma Z}(m_Z))/2$ 
in the expression for $\Delta(Q^2)$, and treat the spread 
$|\Delta_{\gamma Z}(0)-\Delta_{\gamma Z}(m_Z)|/2$ as an estimate of
the NLO uncertainties. This choice keeps all of the experimental
acceptance dependence in the multiplicative correction $1+\delta A(Q^2,y)$, 
properly propagates the bremsstrahlung corrections, 
and reduces the overall size of the theoretical uncertainties. For
experimental kinematics of the E158 experiment, we find 
\beq
\delta A(Q^2,y)=0.006\pm 0.005\ , 
\eeq
where the uncertainty is dominated
by the possible variations of the experimental acceptance and numerical
precision. Employing the 
calculation of Ref.~\cite{cz-marc} for the 
terms $\hat{\kappa}$ and $\rho$, the corresponding contributions from
the $WW$ and $ZZ$ box diagrams, and our definition of 
$\langle \Delta_{\gamma Z}\rangle$ term, the 
residual additive correction in Eq.~(\ref{eq:ALRcorr}) is 
\bea
\Delta(Q^2) &=& \langle \Delta_{\gamma Z}\rangle 
+ \frac{\alpha(m_Z^2)}{4\pi \hat{s}^2} 
-
\frac{3\alpha(m_Z^2)}{32\pi \hat{s}^2(1-\hat{s}^2)}
  (1-4\hat{s}^2)\left[1+(1-4\hat{s}^2)\right] 
= -0.0007\pm 0.0009
\eea
where the uncertainty is dominated by our conservative estimate of the
NLO terms. 

The fact that both corrections $\delta  A(Q^2,y)$ and
$\Delta(Q^2)$ are small is somewhat accidental.  
$\delta A(Q^2,y)$ is dominated by the bremsstrahlung contributions and
the infrared parts of the $\gamma\gamma$ and $\gamma Z$ diagrams that
happen to cancel each other for the asymmetric acceptance of the E158
spectrometer. On the other hand, $\Delta(Q^2)$ is small at the $Q^2$
of E158 due to the cancellation between the $WW$ box diagrams and the
large logarithmic contribution in the $\gamma Z$ box. Such
precise cancellation is not expected at lower momentum transfers, or
for other processes, such as elastic $ep$ scattering. For example, at
$y=0.5$ and $E=12$~GeV, which corresponds to the idealized kinematics for
a proposed \Moller\ scattering
experiments at the Jefferson Lab~\cite{JLab12}, the corrections would
be
\beq
\delta A(Q^2,y=0.5)_{12~\mathrm{GeV}}=-0.024\pm 0.005\ , 
\Delta(Q^2)_{12~\mathrm{GeV}}=-0.0011\pm 0.0010\ .
\eeq
For a 250 GeV fixed target \Moller\ experiment, \eg\ at a future Linear
Collider \cite{MollerLC}, one would find
\beq
\delta A(Q^2,y=0.5)_{250~\mathrm{GeV}}=-0.012\pm 0.005\ , 
\Delta(Q^2)_{12~\mathrm{GeV}}=-0.0002\pm 0.0007\ .
\eeq

\section{Conclusions}

In conclusion, we have computed the QED corrections to the
parity-violating left-right asymmetry \ALR\ in \Moller\ scattering. 
We used a covariant method for removing infrared divergences
without introducing unphysical cutoffs.
For the kinematics of SLAC E158 experiment, the overall corrections
appear to be small, due to a fortuitous
cancellation between electroweak and electromagnetic terms. 
We reduce the theoretical uncertainties due to higher order
logarithmic terms by the appropriate
choice of the couplings used to compute the box diagram
contributions. Our calculation is applicable to a wide range of
fixed target energies, from the proposed \Moller\ scattering
experiments at 12 GeV at the Jefferson Lab~\cite{JLab12}, to the
possible fixed 
target experiments at a future Linear Collider \cite{MollerLC}.

\section{Acknowledgments}

The authors would like to thank Peter Bosted, Stanley Brodsky, Lance
Dixon, Krishna 
Kumar, William Marciano, Michael Peskin, and Frank Petriello for stimulating
discussions. VZ and JS would like to thank SLAC staff for the generous
hospitality during their visits. 
This work has been partially supported by the
National Science Foundation under grant PHY-0140366.

\end{document}